\documentclass[prx,superscriptaddress,aps,amsmath,amssymb,floatfix,reprint,raggedbottom]{revtex4-2}

\usepackage{graphicx}
\usepackage{balance}
\usepackage{xr-hyper}
\usepackage{amsmath}
\usepackage{times}
\usepackage{dcolumn}
\usepackage{xcolor}
\usepackage{lipsum}
\usepackage{braket}
\usepackage{amsfonts}
\usepackage{multirow}
\usepackage[utf8]{inputenc}
\usepackage{url}
\usepackage[ruled]{algorithm2e}
\usepackage[normalem]{ulem}
\usepackage{booktabs}
\usepackage{hyperref}

\DeclareUnicodeCharacter{2009}{\,}
\renewcommand{\url}[1]{}  

\makeatletter
\newcommand*{\addFileDependency}[1]{
  \typeout{(#1)}
  \@addtofilelist{#1}
  \IfFileExists{#1}{}{\typeout{No file #1.}}
}

\makeatother

\makeatletter 
\renewcommand{\fnum@figure}{\textbf{Fig.~\thefigure}}
\makeatother

\usepackage[compact]{titlesec}
\titlespacing{\section}{0pt}{*3}{*2}
\titlespacing{\subsection}{0pt}{*2}{*2}
\titlespacing{\subsubsection}{0pt}{*2}{*2}
\titleformat{\section}{\filcenter\normalfont\small \bfseries}{\thesection.}{1em}{\MakeUppercase}

\emergencystretch=\maxdimen
\begin{document}

\title{
Noise-augmented Chaotic Ising Machines for Combinatorial Optimization and Sampling}
\par

\author{Kyle Lee}\email{kylelee@ucsb.edu}
\affiliation{Department of Electrical and Computer Engineering, University of California, Santa Barbara, Santa Barbara, CA, 93106, USA}
\author{Shuvro Chowdhury}
\affiliation{Department of Electrical and Computer Engineering, University of California, Santa Barbara, Santa Barbara, CA, 93106, USA}
\author{Kerem Y. Camsari}\email{camsari@ece.ucsb.edu}
\affiliation{Department of Electrical and Computer Engineering, University of California, Santa Barbara, Santa Barbara, CA, 93106, USA}

\date{\today}
\begin{abstract}

Ising machines, hardware accelerators for combinatorial optimization and probabilistic sampling problems, have gained significant interest recently. A key element is stochasticity, which enables a wide exploration of configurations, thereby helping avoid local minima. Here, we refine the previously proposed concept of coupled chaotic bits (c-bits) that operate without explicit stochasticity. We show that augmenting chaotic bits with stochasticity enhances performance in combinatorial optimization, achieving algorithmic scaling comparable to probabilistic bits (p-bits). We first demonstrate that c-bits follow the quantum Boltzmann law in a 1D transverse field Ising model. We then show that c-bits exhibit critical dynamics similar to stochastic p-bits in 2D Ising and 3D spin glass models, with promising potential to solve challenging optimization problems. Finally, we propose a noise-augmented version of coupled c-bits via the adaptive parallel tempering algorithm (APT). Our noise-augmented c-bit algorithm outperforms fully deterministic c-bits running versions of the simulated annealing algorithm. Other analog Ising machines with coupled oscillators could {draw inspiration} from the proposed algorithm. Running replicas at constant temperature eliminates the need for global modulation of coupling strengths. Mixing stochasticity with deterministic c-bits creates a powerful hybrid computing scheme that can bring benefits in scaled, asynchronous, and massively parallel hardware implementations.

\end{abstract}
\maketitle

\section{Introduction}
\label{sec:Intro}

\begin{figure*}[!t]
    \centering
s    \includegraphics[width=1 \textwidth]{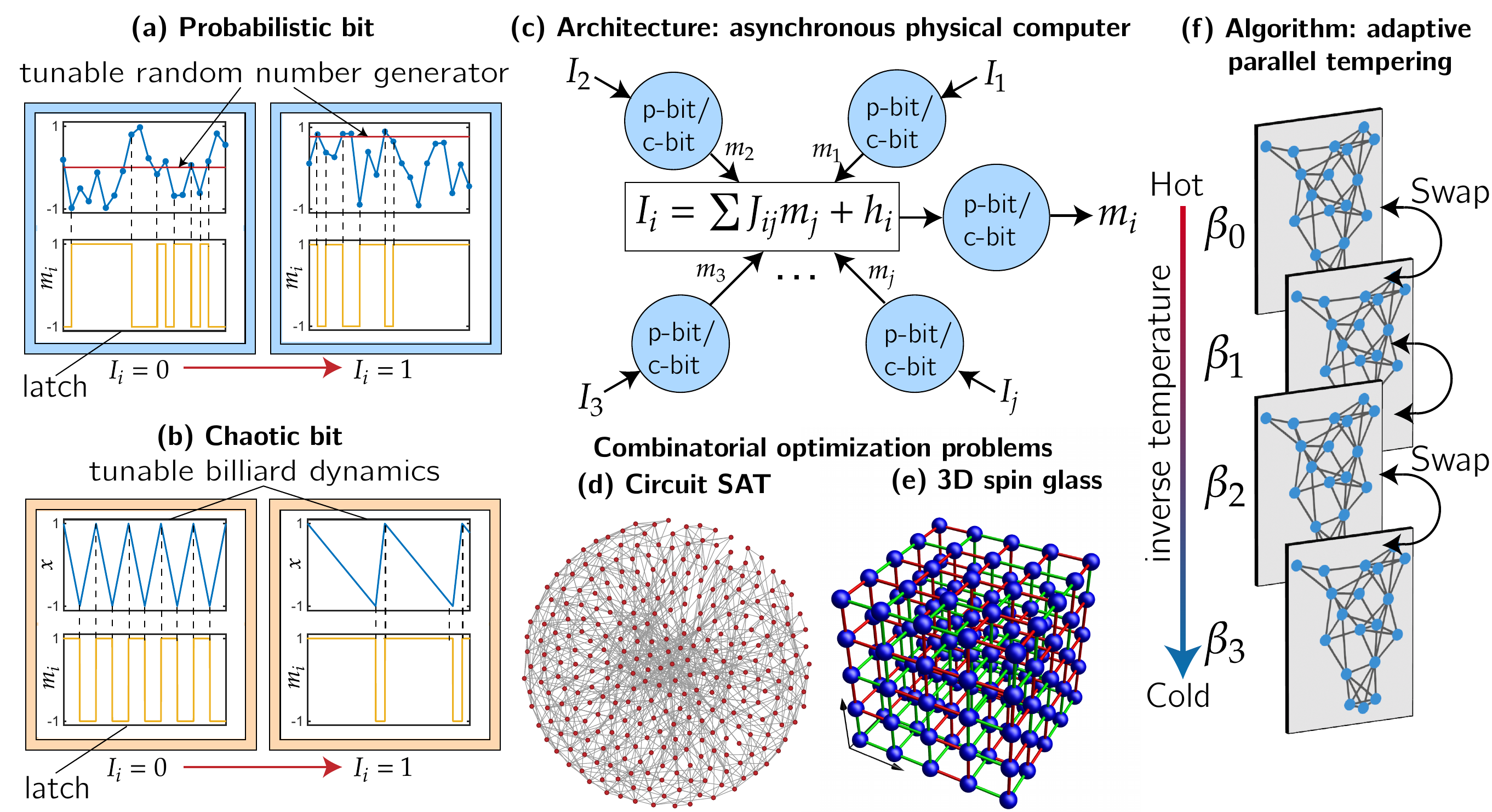}
    \vspace{-15pt}
    \caption{{\footnotesize  (a) A probabilistic bit (p-bit) contains a random number generator with uniform distribution between -1 and 1. $\tanh(I_i)$ is a threshold for latching $m_i$. (b) A chaotic bit (c-bit) features deterministic billiard dynamics with a tunable slope. The billiard is periodic, given that $I_i$ is held constant. When the billiard reaches $-1$ or $+1$, $m_i$ is latched, and the billiard changes direction. (c) p-bits and c-bits form similar, asynchronous architectures, employing the same synaptic function. (d) A 20-bit factorizer conceptualized with p-bits, represented as a graph with 310 nodes and 1200 edges. (e) The 3D spin glass problem, with coupling strengths $J_{ij} \in \{-1,+1\}$. (f) The adaptive parallel tempering (APT) algorithm. Replicas of the same network run at different inverse temperature $\beta$  in parallel. Swaps attempts are made between pairs of replicas with the closest $\beta$, according to the Metropolis criterion (Eq.~\ref{eq:metropolis}).
}}
    \label{fig: fig1}
    \vspace{-7pt}
\end{figure*}

As Moore's law stagnates, there is growing interest in unconventional computing schemes and hardware accelerators. Among these emerging technologies are Ising machines~\cite{mohseni_ising_2022}, employing devices such as probabilistic bits (p-bits)~\cite{camsari_stochastic_2017,borders_integer_2019,kaiser2021probabilistic,aadit_massively_2022}, coupled oscillators~\cite{vadlamani2022equivalence,delacour_mixed-signal_2023,wang2019oim,dutta2021ising,mallick2020using,smith2023polychronous, erementchouk2022computational}, photonic devices~\cite{pierangeli2019large,yamashita_low-rank_2023,litvinenko2023spinwave,inagaki2016coherent}, and other nonlinear elements~\cite{afoakwa2021brim,patel2022logically,hashem2022solving,murali_noise-aided_2023,mallick2023cmos,bae2023ctle,shao2023probabilistic,hizzani2024memristor,lee_correlation-free_2024,si2024energy,fahimi_combinatorial_2021, mahmoodi_versatile_2019}. Ising machines have traditionally focused on providing scaling and prefactor improvements in solving combinatorial optimization problems~\cite{lo2023ising, grimaldi_experimental_2022, singh_hardware_2023}. Efforts to link Ising machines to generative machine learning~\cite{goto2018boltzmann,bohm2022noise,esencan2024combinatorial,laydevant2024training,niazi2024training} and quantum simulation problems are also underway~\cite{chowdhury2023accelerated,chowdhury2023machine,chowdhury2023emulating}. Here, we evaluate and improve an earlier Ising machine concept based on chaotic bits (c-bits), originally proposed in Ref.~\cite{suzuki_chaotic_2013}. Chaotic bits resemble coupled oscillators, following a set of deterministic neuron update rules without explicit randomness. However, unlike oscillator-based Ising Machines that encode the Ising spin in the continuous phase of oscillators, c-bits make use of oscillating billiard balls that set a latch to a $+1$ or $-1$ state. As such, chaotic bits do not require cumbersome subharmonic injection locking schemes to binarize naturally continuous phase variables.

The potential hardware implementation of chaotic Ising machines without any explicit random number generators, while retaining similarity to the mathematics of p-bits, is appealing. However, as we demonstrate in this paper, c-bits without any randomness cannot straightforwardly employ the most powerful Monte Carlo algorithms, such as Adaptive Parallel Tempering, which offer better algorithmic scaling for the Circuit SAT and 3D Spin Glass problems.  We propose and evaluate a \textit{hybrid} computing scheme wherein deterministic c-bits are augmented with stochasticity for improved performance in combinatorial optimization and sampling. The approach we propose may be applied to similar devices, such as certain classes of coupled oscillators, where a continuous global modulation of analog coupling elements may be inconvenient. 

Probabilistic computers perform a discrete Markov Chain Monte Carlo (MCMC) algorithm, simulated in software via Gibbs sampling (FIG.~\ref{fig: fig1}c). p-bit networks are typically described by second-order interactions in energy and stochastic neurons:
\begin{align}
m_i = \text{sgn}[\tanh(\beta I_i) - \text{rand}_u (-1,1)] \label{eq:p-bit neuron}\\
I_i = \sum_{j} J_{ij} m_j + h_i \label{eq:synapse}
\end{align}
where $m_i$$\in$$\{-1,+1\}$ and $\text{rand}_u (-1,1)$ is a random uniform distribution between $-1$ and $1$ (FIG.~\ref{fig: fig1}a, c). $\beta$ is inverse temperature. $J_{ij}$ and $h_i$ represent weights and biases. At equilibrium, p-bit networks sample from the Boltzmann distribution, given by:
\begin{align}
p(\{m\}) = \frac{1}{Z} \exp[-\beta E(\{m\} )] \label{eq:boltzmann}\\ 
E(\{m\}) = - \frac{1}{2} \sum_{i,j} J_{ij} m_i m_j - \sum_{i} h_i m_i \label{eq:energy}
\end{align}
where $\{m\}$ is a spin configuration, $Z$ is the partition function, and $E(\{m\})$ is the energy of a spin configuration. {Chaotic bits follow deterministic update rules, inspired by p-bit equations, which model an oscillating billiard ball and a latch. We reformulate the original chaotic bit equations \cite{suzuki_chaotic_2013,suzuki_monte_2013} to emphasize their similarity to p-bits:}
\begin{equation}
\begin{aligned}
\frac{dx_i}{dt} &= -m_i + \tanh(\beta I_i) \\
x_i = +1 & \implies \text{set } m_i \text{ to } +1 \\
x_i = -1 & \implies \text{set } m_i \text{ to } -1
\label{c-bit neuron}
\end{aligned}
\end{equation}
where $x_i$ represents the position of a billiard between the boundaries $x_i = -1$ and $x_i = +1$. The billiard has a tunable slope, ${dx_i}/{dt}$, which is dependent on the input $I_i$. When the billiard $x_i$ reaches $-1$ or $+1$, the c-bit state $m_i$ is latched to that value, and the billiard changes direction (FIG.~\ref{fig: fig1}b). Note that in numerical or hardware implementations of c-bits, the billiard state $x_i$ in 
{Eq.}~\ref{c-bit neuron} can slightly exceed +1 or $-$1 due to discretization errors.

{The c-bit definition can be motivated from the Boltzmann distribution,  analogous to p-bits and Gibbs sampling. Consider a single spin} {$m_i$} with a \textit{fixed} set of neighbors. For p-bits, combining Eq.~\ref{eq:synapse} {and} Eq.~\ref{eq:boltzmann}, {the conditional probability of} {$m_i$} being $\pm 1$ is expressed as:
\begin{align}
\begin{aligned}
p(m_i = +1) &= \frac{1}{Z} \exp{[\beta (I_i - E_{\text{rest}})]} \\
p(m_i = -1) &= \frac{1}{Z} \exp{[\beta (-I_i - E_{\text{rest}})]}
\end{aligned}
\label{eq:conditional probability with fixed neighbors}
\end{align}
{where} {$Z$} {is the  partition function and} {$E_{\text{rest}}$} {is the part of the global energy (or the environment) that is not dependent on p-bit $i$}. 
{When expressed as a ratio of probabilities,} Eq.~\ref{eq:conditional probability with fixed neighbors} {becomes:}
\begin{align}
\frac{p(m_i=+1)}{p(m_i=-1)} = \exp{(2\beta I_i)}
\label{eq:slope ratio}
\end{align}
{which describes the behavior of a single p-bit given \textit{fixed} neighbors.

As a deterministic unit, c-bit updates do not involve probabilities, but rather they try to achieve the same sampling ratio of $+1$'s to $-1$'s by modifying the lifetime of up and down trajectories of the billiard ball, conditioned on the state of its neighbors (FIG.~}\ref{fig: fig1}{b):}
\begin{equation}
 \frac{\tau(m_i=+1)}{\tau(m_i=-1)}  = \frac{\left|\displaystyle\frac{dx_i}{dt} \quad \mbox{when $m_i=-1$}\right|}{\left|\displaystyle\frac{dx_i}{dt} \quad \mbox{when $m_i=+1$}\right|} = \exp{(2\beta I_i)}
 \label{eq:life}
\end{equation}
where $\tau$ is the lifetime of a c-bit at its current state ($m_i = \pm1$). {In other words, c-bits are defined such that the time average of the c-bit state $\langle m_i \rangle = \tanh{(\beta I_i)}$, analogous to a p-bit.} The key point is that $\tau$ are deterministic, unlike a p-bit whose state is stochastically sampled at each update. 

The comparison of p-bits (Eq.~\ref{eq:p-bit neuron} and Eq.~\ref{eq:synapse}) with c-bits (Eq.~\ref{c-bit neuron} and Eq.~\ref{eq:synapse}) is fundamental. In the case of p-bits, the fundamental theorem of Markov Chains \cite{hespanha2024markov,aarts1989simulated} ensure that a network of p-bits updated sequentially via Gibbs sampling will eventually reach the Boltzmann distribution, defined by Eq.~\ref{eq:boltzmann}. Ultimately, this is what enables mapping problems of interest to p-bit networks and then solving them with algorithms such as simulated annealing and parallel tempering. Networks of c-bits cannot be proved to sample exactly from the Boltzmann distribution due the lack of probabilistic and serial updates.

\begin{figure}[!t]
    \centering
    \includegraphics[width=.90\columnwidth]{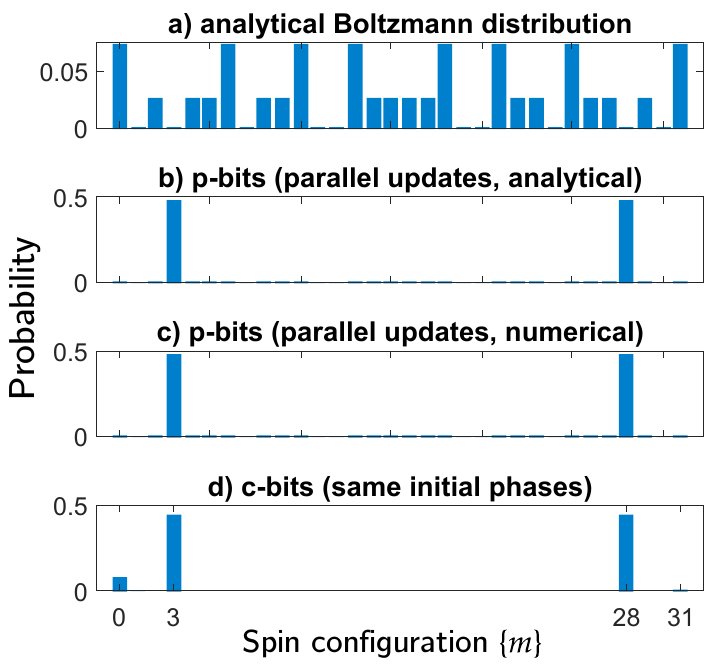}
    \caption{\footnotesize {(a) Distribution of spin configurations for a 5 p-bit full adder, obtained analytically from the Boltzmann distribution. (b) Analytical distribution for p-bits employing parallel updates, as opposed to sequential Gibbs sampling}~\cite{aadit_massively_2022}. {(c) Numerical results for parallel p-bit updates, collected over $10^5$ Monte Carlo sweeps. (d) c-bits exhibit similar pathological behavior when their phases $x_i$ and states $m_i$ are all initialized to the same value, $+1$. Samples are collected over $10^5$ time steps.}
    }
    \label{fig:Parallel Updates Histograms}
\end{figure}

Note the following parallel to p-bits. It is well-known that in Gibbs sampling, simultaneously updating connected nodes leads to pathological oscillations~\cite{koller2009probabilistic,pervaiz_hardware_2017}, preventing easy parallelization of the chain. For c-bits, it seems that the differential equation formulation in continuous time prevents simultaneous updates naturally, as long as the time difference between subsequent billiard ball arrivals is greater than the time it takes to update a latch. When c-bits are discretized in software, latches are set instantaneously within a single time step, $\Delta t$, chosen to solve the differential equations. 

To better understand how critical serial updating is for c-bits, we construct a simple example, with 5 c-bits (or 5 p-bits), interconnected such that their low energy states represent the truth table of a full adder, given by the interactions \cite{smithson2019efficient,pervaiz2018weighted}:

\[
J_{FA} = \begin{bmatrix}
0 & -1 & -1 & 1 & 2 \\
-1 & 0 & -1 & 1 & 2 \\
-1 & -1 & 0 & 1 & 2 \\
1 & 1 & 1 & 0 & -2 \\
2 & 2 & 2 & -2 & 0
\end{bmatrix}
\]
with no bias terms, $h=0$. The columns correspond to carry in, A, B, sum and carry out bits, respectively. FIG.~\ref{fig:Parallel Updates Histograms}a shows the analytically obtained  Boltzmann distribution from Eq.~\ref{eq:boltzmann} at $\beta=0.5$, which can be numerically approximated by \textit{sequential} p-bit updates.

FIG.~\ref{fig:Parallel Updates Histograms}b {shows the probability distribution of 32 possible spin configurations in a 5 p-bit full adder when p-bits are updated in \textit{parallel} rather than sequentially}. The steady-state distribution corresponding to this Markov chain can be obtained analytically \cite{aadit_massively_2022,aarts1989simulated} and describes the numerical sampling (shown in FIG.~\ref{fig:Parallel Updates Histograms}c) well. 

{Remarkably,} FIG.~\ref{fig:Parallel Updates Histograms}d {shows that similar pathological oscillations between states 3 and 28 are observed in c-bit networks when the initial states are synchronized by making all $x_i$ and $m_i$ equal in the beginning. These oscillations go away if the initial phases are sufficiently separate and the integration step, $\Delta$t is small. We conclude that c-bit networks implicitly rely on \textit{sequential} updating for correct operation, despite their continuous time and massively parallel dynamics. While simultaneous c-bit updates in an integration step $\Delta t$ may be considered an artifact of software discretization, hardware implementations
will have similar restrictions due to finite setup/hold times and synapse delays associated with calculating $I_i$.  In Section~\ref{sec:hw}, we briefly discuss implications of this update mechanism for eventual hardware implementations. 

It is important to note that p-bits and c-bits have different notions of time. There is no clear correspondence between a p-bit Monte Carlo sweep (MCS), in which Eqs.~\ref{eq:p-bit neuron} and~\ref{eq:synapse} are sequentially solved for each p-bit in the network, and the time variable $t$ of a c-bit network. In this study, we sample the state $\{m\}$ of a continuously updating c-bit network at natural numbers $t = \{0, 1, 2, \ldots , t_a\}$. {Because of these different notions of time, in order to draw comparisons between p-bit and c-bit algorithms, we shall employ power law and exponential scaling arguments, which hold independent of prefactors. Furthermore, we apply similar algorithmic parameters to both p-bits and c-bits to ensure fair comparisons.}
    
Theoretical work has shown convergence to the Boltzmann distribution for simple toy models {of 2 c-bits}~\cite{suzuki_chaotic_2014}. {However, it has not been determined that larger c-bit networks necessarily sample from the Boltzmann distribution as stochastic p-bits do.} Experimental work suggests that c-bit networks seem to  closely approximate the Boltzmann distribution for 2D Ising models of small sizes~\cite{suzuki_monte_2013}. Furthermore, it has been shown that c-bits match stochastic p-bits for equilibrium statistics on the 2D ferromagnetic Ising model~\cite{suzuki_chaotic_2013} and the Potts model~\cite{suzuki_monte_2013}. c-bits have been used to determine the critical temperature ($T_c$) and critical exponents ($\nu$, $\beta$, and $\gamma$) of the Ising universality class~\cite{suzuki_monte_2013}. Simulated annealing has been demonstrated with c-bits~\cite{suzuki_chaotic_2013, kawashima_fpga_2020}. Additionally, CMOS and analog VLSI implementations have been proposed~\cite{yamaguchi_cmos_2017, yamaguchi_chaotic_2019}. Nonetheless, the theoretical equivalence of p-bits to c-bits without explicit noise is far from clear. Moreover, the initial phase randomization and rounding errors in the solution of ODEs (or physical noise in real systems) might be responsible for this striking correspondence.

Here, we subject c-bits to more stringent tests on three important problem classes, comparing their performance to p-bits. These tests involve a quantum problem, criticality in 3D spin-glasses, and hard combinatorial optimization in the form of integer factorization with invertible logic gates. We conclude that while c-bits  perform similarly to p-bits in algorithms such as simulated annealing, {c-bits also  benefit from added stochasticity, particularly through random swaps used in powerful tempering algorithms.}

Most existing work employs a definition of ${dx_i}/{dt}$ that grows exponentially at cold temperatures ($\beta\gg 1$), with the exception of one previous study~\cite{suzuki_monte_2013}. This is problematic because optimization algorithms such as simulated annealing and adaptive parallel tempering require cold temperatures which may result in unbounded slopes. In this study, we employ a c-bit definition that has a maximum billiard speed of  2 (see Eq.~\ref{c-bit neuron}), suitable for software and hardware implementation. Unless otherwise specified, we employ the Euler's Method with a step of $\Delta t = 0.1$ to simulate the dynamics of coupled c-bits. 

Our bounded c-bit choice is also a potential improvement over previous hardware implementations, which may be suffering from accuracy losses due to exponentially growing slopes. For example, the anomalous increase of energy plots in a prior FPGA implementation  is likely due to the overflows and underflows in slopes (\cite{kawashima_fpga_2020}). 

Finally, in all of our experiments, the initial phases of the billiard balls and the states of the latches are randomized. The asynchronous dynamics of the billiard balls, coupled with this phase randomization ensures nonsimultaneous updates with quasi-random arrivals. In addition, rounding errors and the choice of a naive integration scheme provide additional noise in our simulations. We believe these seem to be the key reasons behind the near-equivalence of c-bits to p-bits, where explicit and implicit inclusion of noise seems to play a central role.

\section{Sampling from the 1D Transverse Field Ising Hamiltonian with c-bits}
\label{sec:Simulated Quantum MC Sampling}

\begin{figure}[!t]
    \centering
    \includegraphics[width=.90\columnwidth]{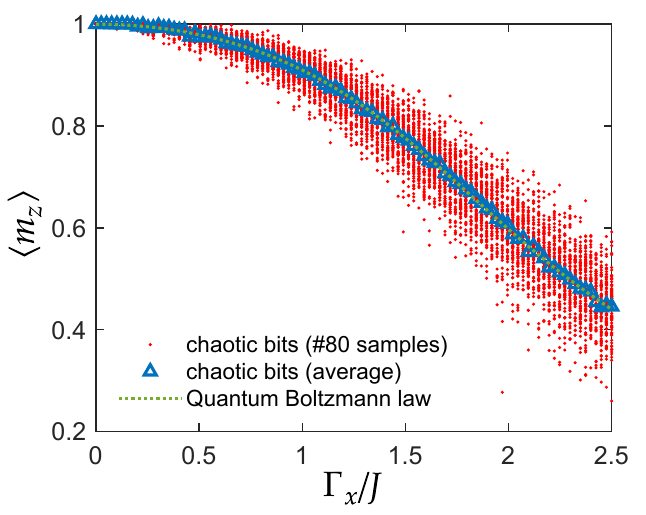}
    \caption{\footnotesize Chaotic bits emulate a 1D ferromagnetic chain ($J_{ij} = +2$) of 8 qubits described by the quantum Transverse Ising Hamiltonian (Eq.~\ref{eq: quantum hamiltonian}). For emulation purposes, we use Suzuki-Trotter decomposition with $R = 250$ replicas. We show average magnetization as a function of the transverse field ($\Gamma_x$) when the system is at constant inverse temperature $\beta = 10$. A symmetry breaking magnetic field of $h_i = 1$ is applied in the $+\hat{z}$ direction such that when $\Gamma_x = 0$, all spins point in the $+\hat{z}$ direction. Red dots show the results of 80 independent trials of c-bit simulations, and a single set of spin configurations is recorded at the end of $t_a = 1000$ time steps, for each trial. Blue triangles show the average numerical result. The green dashed line is an analytical solution from solving Eq.~\ref{eq: quantum boltzmann law} as a function of $\Gamma_x$.
    }
    \label{fig:SQA}
\end{figure}   

As a difficult sampling problem that requires high-quality pseudorandom number generators~\cite{sutton2020autonomous},
we consider a 1D ferromagnetic chain of 8 qubits at constant inverse temperature $\beta = 10$. This system is described by the transverse Ising Hamiltonian in 1D, 
\begin{align}
H_{\text{Q}} = -\sum_{i<j}{J_{ij}\sigma^z_i\sigma^z_j}-\sum_{i}{h_i\sigma^z_i}- \Gamma_x\sum_{i}{\sigma^x_i}
\label{eq: quantum hamiltonian}
\end{align}
where $\sigma_i^x$ and $\sigma_i^z$ are Pauli spin matrices at site $i$. Note that each of these is $2^N \times 2^N$ matrix and is computed as $\sigma_i^{\alpha} = I_2^{\otimes i-1}\otimes\sigma^{\alpha}\otimes I_2^{\otimes N-i}= I_2\otimes\ldots\otimes\sigma^{\alpha}\otimes\ldots\otimes I_2$ with $\alpha\in\{x,z\}$ where $\otimes$ denotes the `Kronecker product', $I_2$ is the $2\times2$ identity matrix, $\sigma^z$ and $\sigma^x$ are Pauli spin matrices (only at the $i$-th term of the product):
\begin{align}
\sigma^z = \left[\begin{array}{cc} 1 & 0\\ 0 & -1\end{array}\right] \quad \text{and} \quad \sigma^x = \left[\begin{array}{cc} 0 & 1\\ 1 & 0\end{array}\right]
\end{align}
The interaction weight $J_{ij} = +2$ for nearest neighbors (with a periodic boundary condition) and $J_{ij} = 0$ otherwise. $h_i = +1$, representing a symmetry breaking magnetic field in the $+\hat{z}$ direction. The Suzuki-Trotter transformation~\cite{suzuki1976relationship, chowdhury2023accelerated} maps the 1D quantum Hamiltonian to the following 2D classical Hamiltonian:

\begin{align}
H_{\text{C}} = -\sum\limits_{k=1}^{R}\left[\sum\limits_{i < j}\cfrac{{J}_{ij}}{R}{m}_{i,k}{m}_{j,k}+\sum\limits_{i}{\cfrac{h_i}{R}\,m_{i,k}}\right.\nonumber \\\left.+\sum\limits_{i}{J}_{\perp }{m}_{i,k}{m}_{i,k+1}\right]
\label{eq:Suzuki Trotter Mapping}
\end{align}
where $J_{\perp}=-{1}/{2\beta}\ln{\tanh{\left({\beta\Gamma_x}/{R}\right)}}$. {$R$ represents the total number of system replicas after applying the quantum-to-classical mapping.} Note the unusual dependence of this  coupling to the inverse temperature, $\beta$. 

For an infinite number of replicas ($R \rightarrow \infty$), this quantum-to-classical mapping is exact. In practice, the error scales as $\mathcal{O}(1/R^2)$ for a given $\beta$. Here, we employ $250$ replicas, thus our system contains $N = 8$ qubits and $R = 250$ replicas, totaling $2000$ classical spins. We simulate our classical system using c-bits. As we vary the value of the transverse field $\Gamma_x$, we measure the average magnetization:
\begin{align}
\langle m_z \rangle = \cfrac{1}{NR}\sum_{k=1}^{R}{\sum_{i=1}^{N}{m_{i,k}}}
\end{align}
In theory, the system should obey the quantum Boltzmann Law:
\begin{align}
\langle m_z\rangle = \cfrac{\text{tr}\left[\exp{(-\beta H_{\text{Q}})}\sum_{i}{\sigma_i^z}\right]}{\text{tr}\left[\exp{(-\beta H_{\text{Q}})}\right]}
\label{eq: quantum boltzmann law}
\end{align}

{$\sum_i \sigma_i^z$ in the numerator is the corresponding quantum operator for measuring spins along the z-direction. Notice that the argument of the `exp\,($\cdot$)' in Eq.~\ref{eq: quantum boltzmann law} is a matrix. Evaluating  this equation becomes intractable for large $N$, as it involves taking the exponential of a $2^N\times2^N$ matrix. When $N$ is small like the system in our example, one can compute exponential of a matrix directly by explicitly writing down $H_Q$ and calculating its exponential. Also, to obtain numerically stable results at low temperatures (high $\beta$), we first diagonalize the Hamiltonian and subtract the ground-state energy from the diagonals which does not affect any observable quantities.}

FIG.~\ref{fig:SQA} shows that numerical results from c-bits show strong agreement with the quantum Boltzmann law, obtained by solving Eq.~\ref{eq: quantum boltzmann law} as a function of $\Gamma_x$. A similar result has been shown in a previous work using stochastic p-bits~\cite{camsari_scalable_2019}. We observe that c-bits appear to show comparable performance to p-bits, despite the fact that c-bits only use explicit randomization for the initialization of spins and phases at the start of each simulation.

\section{Critical Scaling Dynamics: c-bits vs p-bits}
\label{Critical Dynamics}

\begin{figure*}[!t]
    \centering
   \includegraphics[width=1 \textwidth]{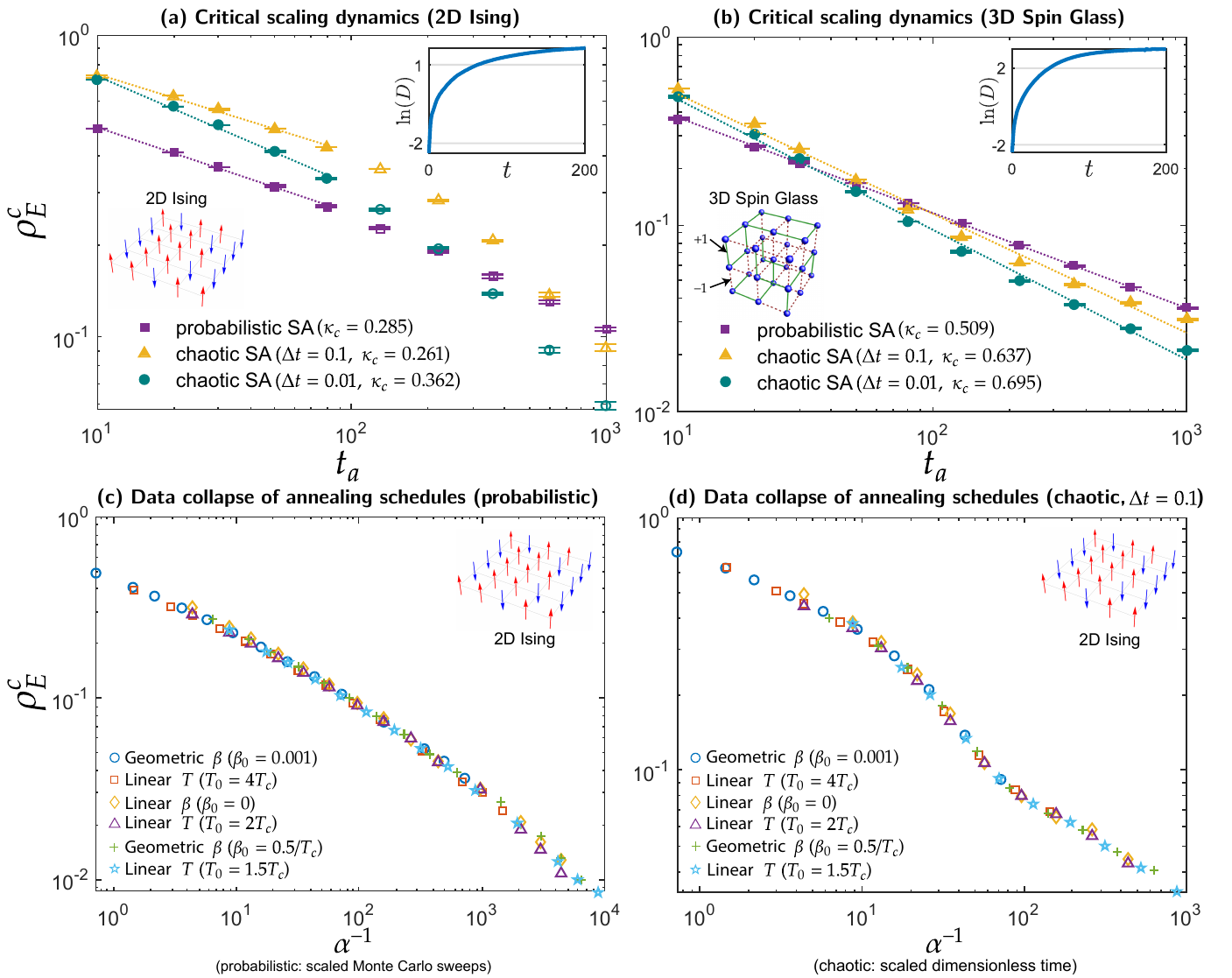}
    \vspace{-15pt}
    \caption{{\footnotesize (a) Residual energy density at the critical point ($\rho_E^c$) as a function of annealing time for a $50 \times 50$ ferromagnetic Ising model lattice with nonperiodic boundary conditions ($J_{ij} = +1$). We perform linear regression on the 5 leftmost points of each data series to avoid finite-size effects. Error bars show $95\%$ bootstrap confidence intervals. {The inset plot shows short-term exponential divergence of c-bit systems with slightly perturbed initial conditions. The slope represents a positive (maximum) Lyapunov exponent, implying chaotic behavior.} (b) $\rho_E^c$ vs. annealing time for cubic spin-glass instances ($L = 11$, $J_{ij} \in \{-1,+1\}$). (c) $\rho_E^c$ as a function of inverse annealing velocity for the 2D ferromagnetic Ising model, using p-bits. Different annealing schedules are shown in a data collapse by considering the quench rate at the critical point, $\alpha$. (d) $\rho_E^c$ vs. inverse annealing velocity for the 2D ferromagnetic Ising model, using c-bits.
}}
    \label{fig: critical scaling dynamics}
    \vspace{-7pt}
\end{figure*}

Next, we discuss critical dynamics of c-bit networks for the 2D Ising model and the 3D spin glass problem. The idea is to quench c-bit networks at varying anneal rates and measure the residual energy as a proxy of topological defects. The density of these defects are qualitatively predicted by the Kibble-Zurek Mechanism (KZM)~\cite{del2014universality}, which relates equilibrium correlation lengths and relaxation times to defect densities borne out of non-equilibrium dynamics. In the present context, the study of KZM allows a comparison of c-bit performance with that of p-bits. In addition, residual energy may be a useful figure of merit for combinatorial optimization where scaling comparisons between different algorithms can be made~\cite{king_quantum_2023}. The residual energy density at the critical point is defined as:
\begin{equation}
\rho_E^c = \frac{\langle E-E_c\rangle}{n}\label{eq:critical residual energy}
\end{equation}
where $E_c$ is equilibrium energy at the critical temperature and $n$ is the number of spins in the system. From the theory of phase transitions, phenomenological scaling arguments can be made to describe the residual energy in terms of an anneal with varying velocities (see Appendix~\ref{sec:appendix1} for details): 
\begin{equation}
\rho_E^c \propto  \alpha^{\kappa_c}
\label{eq:kzMain}
\end{equation}
where $\kappa_c$ is the exponent describing the scaling behavior of the residual energy density at the critical point, and $\alpha$ is the velocity of the anneal measured at the critical temperature for any schedule that starts from some initial temperature and ends at the critical temperature:
\begin{equation}
\alpha = \left.\frac{d  \lambda(t)}{dt}\right|_{t=t_a}
\label{eq:alphafactorM}
\end{equation}
One way to make contact with known critical exponents and our numerically calculated exponents is to measure the number of defects via kinks~\cite{bando2020probing} in Ising models, but our purpose here is to provide a comparative analysis with p-bits and c-bits, so direct verification of the exponents is not necessary.

First, we consider a $50 \times 50$ ferromagnetic Ising model lattice with nonperiodic boundary conditions ($J_{ij} = +1$). For a given annealing time $t_a$, $\beta$ is geometrically increased from $\beta = 0.001$ to the critical temperature, $\beta ={T_c}^{-1}$. For this problem, we use the exact critical temperature calculated at the thermodynamic limit~\cite{PhysRev.65.117} as an approximate $T_c$ for the finite-size model:  $T_c = {2}/{\ln(1+\sqrt{2})} \approx 2.269$). When annealing terminates at the critical point, the energy of the spin configuration is recorded to calculate $\rho_E^c$. We average over 3000 randomized trials.

Although there is no obvious correspondence between a Monte Carlo sweep for behavioral models of p-bits and the time step of a c-bit,  power-law exponents can be compared objectively since they do not depend on prefactors. FIG.~\ref{fig: critical scaling dynamics}a shows that c-bits and p-bits demonstrate similar critical scaling dynamics on the 2D ferromagnetic Ising model. Linear regression is performed on the 5 leftmost points in order to avoid finite-size effects. Both p-bit and c-bit curves exhibit downward concavity, but different time scales mean that this concavity does not manifest at the same time $t_a$ for p-bits and c-bits. 
We provide the results of using Euler's method in steps of $\Delta t = 0.1$ and $\Delta t = 0.01$. The c-bit curve using $\Delta t = 0.01$ appears to follow a similar power law scaling, but with a steeper slope $|\kappa_c|$. The reason for this discrepancy is not clear. On the one hand, smaller $\Delta t$ implies higher theoretical solver precision, while on the other hand, smaller $\Delta t$ leads to more round-off errors due to double precision. It must also be noted that a lower $\Delta t$ requires more computational effort, and in hardware, $\Delta t$ may have a lower bound dictated by the physical latch required to set the state of the oscillator. 

FIG.~\ref{fig: critical scaling dynamics}b shows a similar experiment conducted on select instances of the 3D spin glass problem, studied by D-Wave~\cite{king_quantum_2023}. We consider 300 different spin glass instances on a cubic lattice with side length $L = 11$ and $J_{ij} \in \{-1,+1\}$. Analogous to FIG.~\ref{fig: critical scaling dynamics}a, for a given annealing time $t_a$, we geometrically increase $\beta$ to the inverse critical temperature $1/T_c$. For 3D spin glasses, $T_c$ is approximately calculated as $\approx 1.1$~\cite{baity-jesi_critical_2013}). For each of 300 problem instances, we take an ensemble average over 40 random seeds. We again observe similar power-law scaling. {Our p-bit results match the numerical results in the literature, with $\kappa_c \approx 0.51$~\cite{king_quantum_2023}}. However, the values of $\kappa_c$ do not seem to be similar when comparing between p-bits and c-bits. The $\Delta t$ dependence of c-bit slopes also is an indication that the underlying dynamics for c-bits might depend on separating near arrivals that latch a state (where Eq.~\ref{eq:synapse} is always assumed to be infinitely fast). We conclude that while c-bits seem to obey a similar power law as p-bits, their dynamics seem different in subtle ways. 

One feature of chaotic dynamics is  sensitivity to initial conditions. This can be  tested by calculating Lyapunov exponents which measure the distance between a chosen initial condition and a slightly perturbed version of it as a function of time. If the distance in phase space grows exponentially in time, this is an indication of chaotic dynamics. In general, there is a spectrum of Lyapunov exponents, but the dynamics are dominated by the maximum Lyapunov exponent, which can be empirically measured.  In the {inset plots of} FIG.~\ref{fig: critical scaling dynamics}a and FIG.~\ref{fig: critical scaling dynamics}b, {we show maximum Lyapunov exponents for the 2D Ising model and the 3D spin glass problem, respectively. We observe that in both cases the log distance acquires a linear slope and then saturates. The linear and positive slope exhibits chaotic dynamics before saturating. The saturation of the distance is indicative of the finite size of the phase space where a distance between two trajectories has an upper bound. 

The specifics of the calculation are as follows: the phases $\mathbf{x}$ and the states $\mathbf{m}$ are randomly initialized. We then introduce a perturbation in the initial conditions by offsetting the phase $x_i$ of a randomly selected c-bit by a magnitude of $0.1$. We then consider the Euclidean distance between the phases of the perturbed and unperturbed systems: $D = ||\mathbf{x} - \mathbf{x}'||$, where $\mathbf{x}'$ represents the perturbed system's phases. The log Euclidean distance is plotted over 200 time steps, with the system at constant inverse temperature $\beta = 1$. The slope represents the Lyapunov exponent $\lambda$, as $D \sim \exp{\lambda t}$. The results are averaged over 100 trials, where each trial compares one unperturbed system $\mathbf{x}$ with 10 perturbed versions, $\mathbf{x}'$}.

In FIG.~\ref{fig: critical scaling dynamics}c-d, we show how both c-bits and p-bits exhibit a clear power-law relationship with critical exponents  by comparing different annealing schedules with different initial temperatures. As Eq.~\ref{eq:kzMain} predicts, the simulations should exhibit universal behavior if plotted as a function of how fast the anneal is performed near the critical point. When plotted as function of annealing velocity, $\alpha$ (Eq.~\ref{eq:alphafactorM}), all c-bit and p-bit plots fall on top of each other, demonstrating an excellent collapse. The different exponents of c-bits and p-bits and the time-step dependence of the c-bit networks indicate that c-bit networks may not be a drop-in replacement for p-bit networks whose steady-state provably takes samples from the Boltzmann distribution~\cite{aarts1989simulated}.

\section{Adaptive Parallel Tempering}
\label{APT}

\begin{algorithm}
\caption{Adaptive Parallel Tempering with p-bits or c-bits}\label{alg:APT}

    \KwIn{weights $J_{ij}$, biases $h_i$, number of replicas $N$, inverse temperature profile $\beta$, time steps per swap, simulation time $t_a$}
    \KwOut{Spin states $\{m\}$ corresponding to the minimum energy $E_0$}
    
    Initialize $N$ system replicas to random spins\;

    \For{time $t = 0$ to $t_a$}{
        \For{$N$ replicas at different $\beta$}{
            Simulate p-bit Monte Carlo sweeps using Eq.~\ref{eq:p-bit neuron} and~\ref{eq:synapse}, or c-bit dynamics using Eq.~\ref{eq:synapse} and~\ref{c-bit neuron}\;
        }
        
        \If{time $t$ is a multiple of time steps per swap}{
            \tcp{Probabilistic swap attempt}
            \eIf{it is an even numbered swap attempt}{
                Select (even, odd) replica pairs\;
            }{
                Select (odd, even) replica pairs\;
            }
    
            \For{each replica}{
                Compute energy $E\{m\}$ (Eq.~\ref{eq:energy})\;
            }
    
            \For{each pair of adjacent replicas}{
                \If{the Metropolis criterion is met (Eq.~\ref{eq:metropolis})}{
                    Swap the spin state $\{m\}$ between replicas\;
                    \tcp{For c-bits, also swap the billiard state $\{x\}$ between replicas}
                }
            }
        }
    }
\end{algorithm}

Critical scaling dynamics are intriguing from a physics perspective, but it is not essential that c-bits obey the same physics as p-bits in an optimization context. For optimization, we use an adaptive version of the parallel tempering algorithm, abbreviated as APT~\cite{isakov2015optimised,mohseni2021nonequilibrium,nikhar_all--all_2023}. A given p-bit or c-bit network is duplicated into $N$ system replicas, with each replica at a different inverse temperature $\beta$ (FIG.~\ref{fig: fig1}f). In fixed time intervals (time steps per swap attempt), a probabilistic swap attempt is conducted between adjacent replica pairs according to the Metropolis criterion:
\begin{align}
P(\text{swap}) = \min[1, \exp(- \Delta \beta \Delta E)]
\label{eq:metropolis}
\end{align}

If this condition is met, then the adjacent replicas swap their spin states $\{m\}$. Alternatively, replicas may swap their temperature values with a rearrangement of replica indices, which may be more convenient in dedicated hardware implementations. Intuitively, hot replicas (small $\beta$) explore the state space of possible spin states $\{m\}$. The Metropolis criterion makes low energy configurations more likely to swap to cold replicas (high $\beta$). Over time, the coldest system replica tends toward the ground energy, $E_0$.

The APT algorithm can be applied using either p-bits or c-bits. While the neuron update rule is stochastic for p-bits (Eq.~\ref{eq:p-bit neuron}), it is deterministic for c-bits (Eq.~\ref{c-bit neuron}). Replica swaps via the Metropolis criterion are always probabilistic (Eq.~\ref{eq:metropolis}). Therefore, while p-bit APT is a purely probabilistic algorithm, c-bit APT is a hybrid chaotic-probabilistic scheme, where neuron updates are deterministic and replica swap attempts are probabilistic. In typical p-bit APT, the great majority of the random numbers are used for neuron updates rather than replica swap attempts. {In contrast, c-bit APT requires orders of magnitude fewer random numbers.} However, this potential benefit critically depends on the hardware cost of implementing c-bits, which may be greater than that of p-bits. Given the relative ease of creating pseudo random number generators through compact linear-feedback-shift-register circuits~\cite{singh2024cmos}, we suspect that the c-bits may surreptitiously contain PRNG-like circuits in their implementation. {While p-bits and c-bits appear to be similar in their computational power, it is important to note that only p-bits enjoy a mathematical certainty that they sample from the Boltzmann distribution at steady state.} We further discuss these ideas in Section~\ref{sec:hw}. 

The algorithm is referred to as `adaptive' because we employ an instance-specific pre-processing method that finds a suitable $\beta$ schedule for a given problem. As shown in FIG.~\ref{fig:factorization}e, this process ensures that replica-swap probabilities are nearly uniform, avoiding bottlenecks during the exchange process (see Appendix~\ref{sec:appendix2}). The complete APT algorithm is summarized in pseudo code (Algorithm~\ref{alg:APT}).

\section{Optimization of the 3D Spin Glass Problem}
\label{Spin Glass}

\begin{figure}[!t]
    \centering
    \includegraphics[width=.90\columnwidth]{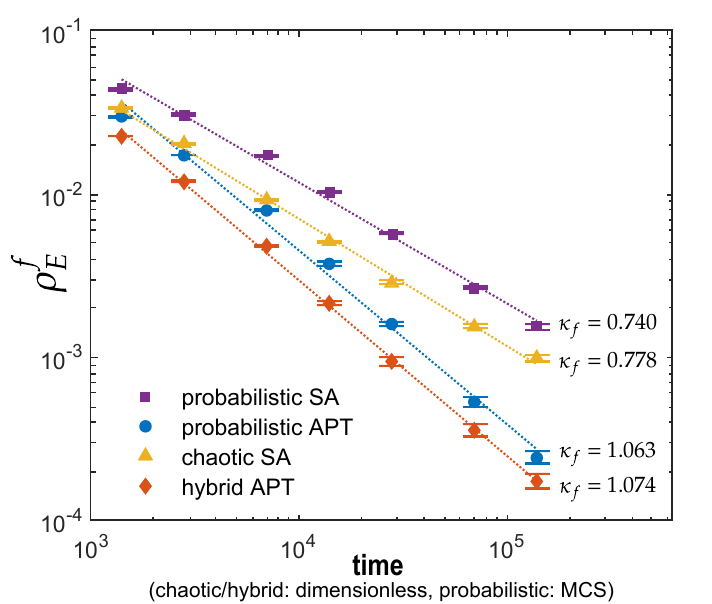}
    \caption{\footnotesize Residual energy density ($\rho_E^f$) as a function of  total computational time ($Nt_a$) for cubic spin-glass instances with side length $L = 7$ ($J_{ij} \in \{-1,+1\}$). Curves scale as a power law: $\rho_E^f \propto t_a^{-\kappa_f}$. Error bars show 95\% bootstrap confidence intervals.}     \label{fig:spinglassoptimization}
\end{figure}

We consider 300 spin glass instances on a cubic lattice with side length $L = 7$ and $J_{ij} \in \{-1,+1\}$, borrowed from Ref.~\cite{king_quantum_2023}. Purely probabilistic APT (p-bits) and hybrid chaotic-probabilistic APT (c-bits) are performed using similar parameters. We conduct APT pre-processing on 1 spin glass instance, returning a $\beta$ schedule for 14 system replicas. In principle, to achieve the best performance, one could apply the APT pre-processing to each of 300 different problem instances and obtain 300 unique $\beta$ schedules. Since this is a comparative study between p-bits and c-bits, we forego this step. Tuning parameters is not of particular concern.

For a given simulation time $t_a$, each of 300 spin glass instances is optimized using APT. For each simulation, we record the lowest energy $E$ sampled across all replicas. We introduce residual energy as an optimization performance metric, defined as:
\begin{align}
\rho_E^f = \frac{\langle E-E_0\rangle}{n}\label{eq:residual energy}
\end{align}
where $E_0$ is the ground state energy and $n$ is the number of spins in the system. FIG.~\ref{fig:spinglassoptimization} shows $\rho_E^f$ vs. computational time, where the time axis accounts for the total time steps summed across all $N = 14$ replicas that run in parallel. We take an ensemble average over 40 random seeds that randomize initial states (for p-bits and c-bits) and phases (for c-bits). 

We devise a similar experiment for simulated annealing. In order to draw a comparison to APT, simulated annealing is conducted 14 times in parallel, with $\beta$ increasing linearly from 0 to 10 during simulation time $t_a$. We record the lowest energy sampled across all replicas to compute the residual energy $\rho_E^f$. We take an ensemble average over 40 random seeds.

To compare p-bit and c-bit performance, we employ similar power-law scaling arguments as we did for critical scaling dynamics. $\rho_E^f = A t_a^{-\kappa_f}$, where $A$ is a constant accounting for the difference in p-bit and c-bit prefactors, and $\kappa_f$ is the power law scaling exponent. FIG.~\ref{fig:spinglassoptimization} shows that p-bit SA exhibits a similar scaling exponent $\kappa_f$ to c-bit SA, while purely probabilistic APT yields a similar $\kappa_f$ as hybrid probabilistic-chaotic APT. Furthermore, the APT algorithms show $\kappa_f$ of larger magnitude than the SA algorithms, indicating a faster convergence to the ground energy $E_0$. Two conclusions can be drawn from our results. First, in the optimization setting, there is a a remarkable similarity between c-bits and p-bits since they essentially show the same scaling for both SA and APT. Second, the hybrid APT algorithm we propose for c-bits shows superior performance over its fully deterministic counterpart.
Besides better algorithmic scaling, the APT algorithm we propose enjoys another benefit: because coupling strengths are not adjusted globally, overflow or underflow issues related to the lifetime of c-bits become less of a concern, and they are a one-time problem to solve (for instance, past definitions of the c-bit have $dx_i/dt$ that grow exponentially as $\beta$ increases, thus overflows may occur at cold temperatures~\cite{suzuki_chaotic_2013}).

The hybrid chaotic-probabilistic APT algorithm we propose here {may be applicable to other approaches, such as the noise-injected oscillator-based Ising machines proposed in Ref.~\cite{bohm2022noise}, that are designed to enable sampling from the Boltzmann distribution}. However, the simplicity of the c-bit and its state-based (rather than phase) representation through explicit latches may still be more appealing. 


\section{Adaptive Parallel Tempering for Semiprime Factorization}
\label{Factoring}

\begin{figure*}[!t]
    \centering
    \includegraphics[width=0.95 \textwidth]{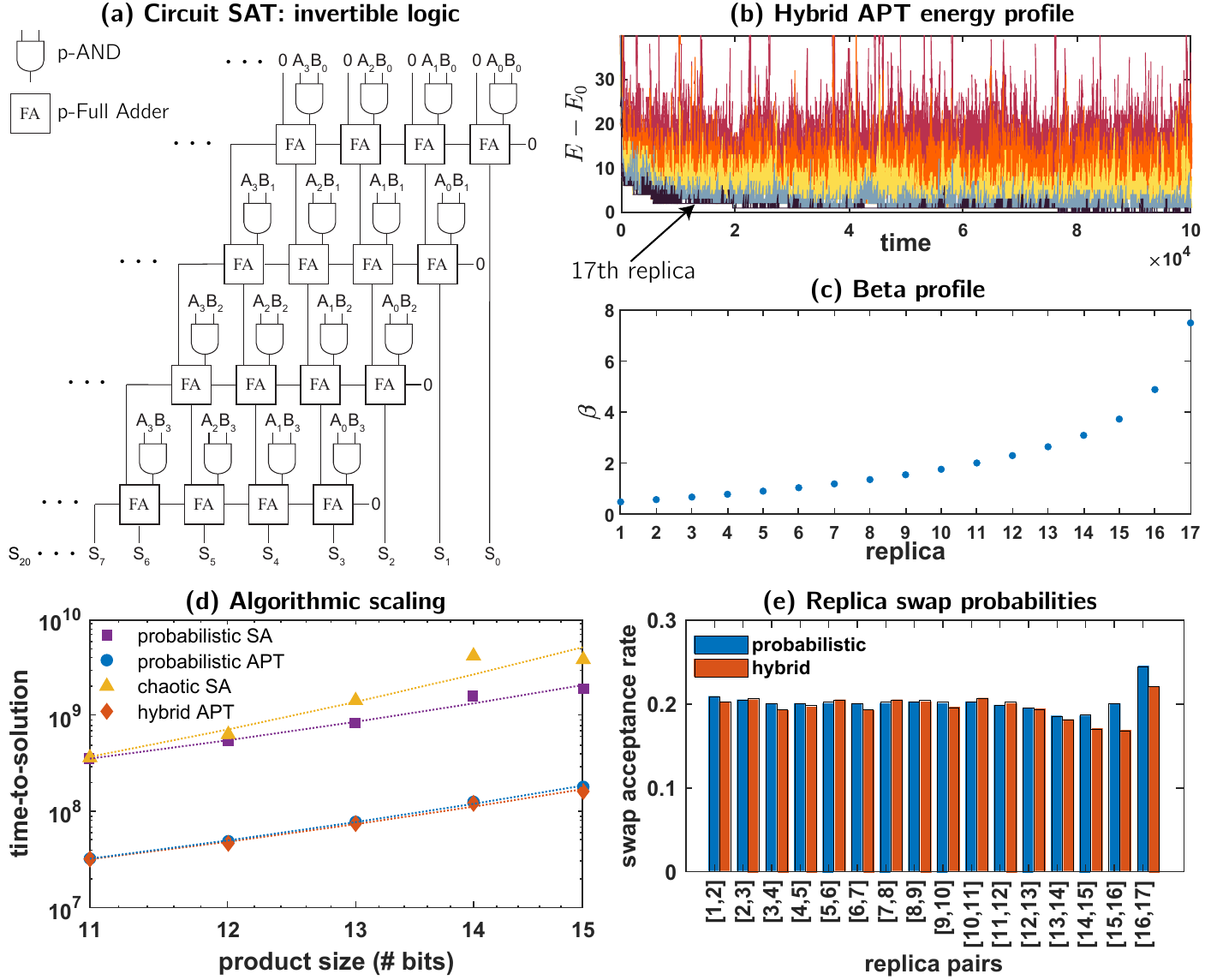}
    \vspace{-10pt}
    \caption{{\footnotesize  (a) p-bits are utilized to construct AND gates and full adders, that are the fundamental gates for a multiplier. The product bits are clamped to a semiprime number as in  invertible boolean logic~\cite{camsari_stochastic_2017}. (b) Energy profile of the 5 coldest replicas during hybrid chaotic-probabilistic APT. $E - E_0$ represents distance from ground state energy. (c) $\beta$ profile obtained from APT preprocessing (see Appendix~\ref{sec:appendix2}). (d) Time-to-solution vs. product size using a 10-bit $\times$ 10-bit factorizer for semiprime numbers. p-bit SA yields a slope of 0.441, while p-bit APT yields a slope of 0.438. c-bit SA exhibits a slope of 0.658, and hybrid APT has a slope of 0.420. (e) Average replica swap probabilities from factoring a specific semiprime number over 100 trials.
}}
    \label{fig:factorization}
    \vspace{-7pt}
\end{figure*}

As a final example, we consider a 10-bit $\times$ 10-bit invertible multiplier circuit composed of p-AND gates and p-Full Adders~\cite{camsari_stochastic_2017,aadit_massively_2022} (FIG.~\ref{fig:factorization}a). This multiplier is represented by an undirected network of $n = 310$ p-bits with 1200 weighted connections, normalized such that $J_{ij} \in [-1,1]$. Of the $310$ spins in the system, 20 specific spins represent the product bits, while another set of 20 specific spins represents each of the two 10-bit factors. Using bipolar spin states ($m_i \in{\{-1,+1\}}$), traditional binary $0$s are instead represented as negative spins $m_i = -1$. Clamping the multiplier's 20-bit output to a semiprime number configures the system such that its ground state, $E_0$, solves for the correct prime factors. To accommodate semiprime numbers that are smaller than 20 bits, we pad the most significant bits with zeros. This scheme is referred to as invertible Boolean logic, in which p-bit based logic gates are run in reverse, similar in spirit to those used in memcomputing.~\cite{camsari_stochastic_2017,manukian2017memcomputing}. 

We employ identical parameters for both probabilistic APT and hybrid chaotic-probabilistic APT. Our APT pre-processing method finds a suitable $\beta$ profile (FIG.~\ref{fig:factorization}c). We consider $N = 17$ system replicas, each running at a different inverse temperature $\beta$ for a duration of $t_a = 10^5$ time steps. Probabilistic swap attempts are made between adjacent replica pairs once every 10 time steps, according to the Metropolis criterion (Equation~\ref{eq:metropolis}). These parameters yield $Nt_a = 17 \times 10^5$ time steps of total computation summed across replicas. 

During an APT simulation of $t_a = 10^5$ time steps, if the coldest replica (highest $\beta$) detects the correct factors at any time step, the simulation is considered a successful solve (FIG.~\ref{fig:factorization}b). Only the coldest replica is considered to remove the effect of random search statistics which may play a role in small problem sizes.  For our comparative study, optimizing the algorithms' performance is not necessary.

We adopt a commonly used time-to-solution (TTS) metric~\cite{kowalsky20223}:
\begin{align}
\langle \text{TTS} \rangle = (Nt_a)\left\langle{ \frac{\ln{(1 - 0.99)}}{\ln{[1-p(Nt_a)]}} }\right\rangle
\label{eq:TTS}
\end{align}

where $Nt_a$ is the total computational time summed across all replicas. The time-to-solution represents the amount of computation necessary to factor semiprimes with a 99\% success probability (see Appendix~\ref{sec:appendix3}). For a given product size (number of bits), we consider 180 unique semiprime numbers. We attempt to factor each semiprime number 80 times. $p(Nt_a)$ is measured as the number of successful solves divided by the total number of attempted factorizations.

We devise a similar scheme for simulated annealing for the sake of comparison. We anneal 17 system replicas in parallel. Each replica runs for $t_a = 10^5$ time steps, and $\beta$ is linearly increased from 0 to 10. In order to avoid random-find effects due to small problem sizes, we only check for the correct factors at the last 1000 time steps. If any of the 17 replicas finds the correct factors, the trial is considered a successful solve. For a given problem size (\# bits), we again consider 180 unique semiprimes. We attempt to factor each semiprime 120 times. $p(Nt_a)$ and time-to-solution are calculated in a fashion analogous to APT. 

Once again, in order to rule out implementation-dependent differences in unit time between c-bits and p-bits, we perform a scaling analysis, where TTS vs. problem size is assumed to have exponential dependence: TTS $ = A \exp{(kx)}$, where $x$ is the product size in bits, while $A$ and $k$ are constants.

Using this method of comparison, FIG.~\ref{fig:factorization}d shows that purely probabilistic APT exhibits remarkably similar algorithmic scaling to the newly introduced chaotic-probabilistic hybrid scheme. Moreover, they both exhibit better scaling than SA-based approaches. Pure chaotic SA, which has been commonly implemented~\cite{suzuki_chaotic_2013}, shows the worst scaling.  Furthermore, both p-bits and c-bits observe a boundary effect  as the problem size approaches 20 bits, where the time-to-solution actually decreases (for all the algorithms). This is a feature particular to the fixed 20-bit factorizer circuit: since numbers larger than 20 bits cannot be represented by the circuit, the solution space is reduced as we approach 20 bits. We exclude this data because it is uninformative for algorithmic scaling. 

The two computing schemes not only share similar algorithmic scaling but also exhibit similarities in their underlying mechanisms. We consider a specific 11-bit semiprime number and conduct APT for $t_a = 10^5$ time steps. We average over 100 random seeds. FIG.~\ref{fig:factorization}e shows the probability of the Metropolis criterion dictating a swap between a given pair of replicas. Remarkably, the hybrid APT scheme follows a similar qualitative distribution as the probabilistic APT scheme.

The choice of integer factorization as a benchmark problem is simply due to the challenging combinatorial nature it poses to stochastic local search algorithms~\cite{mosca2022factoring}. Specialized factoring algorithms starting with trial division and more complex field sieve approaches exhibit superior algorithmic scaling. It is important to note, however, that the factoring representation we use in this work is based on the more general NP-hard Circuit SAT problem, for which no generic algorithmic speed up is known. As such, searching for scaling and prefactor improvements by dedicated hardware is highly desired.

\section{Possible Hardware Realizations}
\label{sec:hw}
Based on the original c-bit definition that is different the one we used in this paper, chaotic bit implementations have been made in digital and analog CMOS~\cite{yamaguchi_chaotic_2019, yamaguchi_cmos_2017,yoshioka2023fpga}. 
We now make some remarks regarding hardware implementations of chaotic networks, however, a concrete discussion of hardware implementations is beyond our scope. 

As a first evaluation of the algorithm we proposed, the results we presented are \textit{hardware-agnostic} based on scaling laws and algorithmic exponents. However, there are restrictions that would be encountered in any hardware implementation. 

The first point is related to the synapse calculation time. In our software implementation, the synapse (Eq.~\ref{eq:synapse}) is updated every time step, so c-bits states are always updated using up-to-date information from their neighbors. Just as in asynchronous networks of p-bits \cite{pervaiz_hardware_2017}, the asynchronous operation  of c-bits require a fast local field calculation.  This  places stringent requirements on the network topology since dense networks will require more time to update the synapse. 

The second point is related to the latch mechanism: as we showed in FIG.~\ref{fig:Parallel Updates Histograms}, if the phases of two different billiards are too close to each in other in time, the latches might erroneously update at the same time depending on the way the latch is implemented, especially in dense network topologies.

Beyond digital implementations, c-bits can make use of ``natural'' latches. For example, the physics of spin-torque switching of a non-volatile nanomagnet~\cite{sun2000spin} can function as the latch: the magnet will switch only if a current threshold is reached, and the magnet will exhibit hysteresis until the current threshold changes sign, much like the billiard-ball metaphor used in c-bits. 

A purported benefit of c-bits is the absence of explicit random numbers in their update rules. However, {it is not clear whether the cost of c-bit hardware implementation is significantly cheaper than that of p-bits}. First, low-cost pseudo random number generators, such as LFSRs, do not significantly increase the complexity of implementing a p-bit. For example, an $n$-bit LFSR requires about $32\times n$ transistors to implement~\cite{singh2024cmos}. In contrast, constructing a c-bit digitally may be significantly more complex than this cost due to the necessity of an oscillator with a tunable duty cycle. Additionally, there may be ``hidden'' sources of randomness, such as floating point round-off errors in simulation or thermal noise in physical implementation.

Augmenting c-bits with explicit randomness in powerful tempering algorithms such as the one considered in this paper could offer further benefits. We believe that while the lack of random numbers may be an important c-bit advantage, comparisons must be made in concrete implementations in order to be meaningful (for example, against the most energy-efficient and compact hardware solutions of p-bits, using magnetic nanodevices~\cite{borders_integer_2019}). {We refrain from making preemptive assumptions about the scalability and efficiency of c-bit hardware, as specific implementation details are unknown. p-bit Monte Carlo sweeps are not 1-to-1 with c-bit's time variable $t$, so speculations are not particularly informative until implementation.} Nevertheless, we believe that the similar algorithmic scaling performance of c-bits and p-bits across the wide variety of problems considered in this paper is very promising, providing valuable insights into the level of stochasticity required for effective probabilistic computing.
\section*{Conclusion}
\label{sec:conclusions}

In this work, we evaluated and improved deterministic chaotic bit networks that combine billiard dynamics with latches. {We note that while algorithm parameters were not excessively tuned, using similar parameters for p-bit and c-bit versions of SA and APT enabled fair comparison of algorithmic scaling performance. We demonstrated that augmenting deterministic c-bits with stochasticity via probabilistic APT replica swaps exhibits algorithmic scaling comparable to APT using fully stochastic networks of p-bits.} Additionally, noise-injected c-bit APT achieved better algorithmic scaling than common simulated annealing-based algorithms. Notably, the proposed adaptive parallel tempering algorithm is applicable to {certain classes of oscillator Ising machines and similar nonlinear devices.}

Moreover, we have shown that c-bits approximately sample from the quantum Boltzmann law in a 1D TFIM model. On the 2D Ising and 3D spin glass problems, c-bits qualitatively exhibit similar critical scaling dynamics as p-bits, though a precise correspondence between their measured scaling exponents $\kappa_c$ is unclear. While intriguing from a physics perspective, these critical dynamics are not essential to optimization, for which c-bits show strong promise. Combining stochastic p-bits and deterministic c-bits in asynchronous and massively parallel hardware implementations could create powerful domain-specific computers for combinatorial optimization and probabilistic sampling. 

\section*{Methods}

Numerical simulation uses double precision (64-bit) in C++ programming language. Random numbers are generated using the Mersenne Twister engine. Spins (and initial billard states, $x_i$, for c-bits) are randomly initialized at the beginning of every SA and APT simulation. For c-bits, we employ Euler's Method with a step of $\Delta t = 0.1$ unless otherwise stated. At a given step, if the billiard $x_i$ is greater than or equal to $+1$ or less than or equal to $-1$, then $m_i$ is latched to that value. 

For the 2D ferromagnetic Ising model, we obtain the critical energy experimentally. Using p-bits, we conduct $10^6$ Monte Carlo sweeps at constant critical temperature $T_c = {2}/{\ln(1+\sqrt{2})}$. We record the average energy over the last $10^5$ sweeps. We take an ensemble average over 40 random seeds.

For the 3D spin glass instances, using p-bits, we conduct $10^6$ Monte Carlo sweeps at constant critical temperature $T_c = 1.1$, a value found from a previous study~\cite{baity-jesi_critical_2013}. For each of 300 problem instances, we record the average energy over the last $10^5$ sweeps. 

\appendix

\subsection{Appendix 1}
\label{sec:appendix1}

In this Appendix, we describe the details of the Kibble-Zurek Mechanism (KZM)  results shown in FIG.~\ref{fig: critical scaling dynamics}. Although we closely follow the arguments presented in Ref.~\cite{del2014universality}, our purpose is to provide precise details about our own measurement procedure. 

According to the scaling hypothesis in phase transitions, macroscopic quantities such as relaxation times and correlation lengths diverge at the critical point. These quantities typically exhibit a power-law relationship with respect to a tuning parameter, $\lambda$. In our context, $\lambda$ is the temperature, $T$, or its inverse, $\beta$. The power-law dependence of key quantities with respect to $\lambda$ around the critical point can usually be justified by notions of scale-invariance and self-similarity:
\begin{equation}
\xi(\lambda)=\frac{\xi_0}{|\lambda_c - \lambda|^{\nu}}
\label{eq:lengthcorr}
\end{equation}
where $\xi$ is the correlation length, $\nu$ is the critical exponent for correlation lengths,  and 
\begin{equation}
    \tau(\lambda)=\frac{\tau_0}{|\lambda_c -\lambda|^{z \nu}}
    \label{eq:taucorr}
\end{equation}
where $\tau$ is the relaxation time and $z \nu$ is the critical exponent for the relaxation time.  Consider now a general annealing schedule, $\lambda(t)$ starting from an initial, $\lambda(t=0)=\lambda_i$ and ending at the critical $\lambda(t=t_a)=\lambda_c$. Since the key quantities of interest diverge near the critical point, the system remains roughly in equilibrium until it approaches the critical point. The non-equilibrium dynamics of interest occur close to the critical point, thus any annealing schedule $\lambda(t)$ can be linearized around $\lambda_c$, which is reached at $t=t_a$:
\begin{equation}
\lambda(t) = \lambda_c + \alpha (t - t_a)  + O[(t-t_a)^2]
\label{eq:linanneal}
\end{equation}
where $\alpha$ is a key quantity representing the velocity of the anneal as the system approaches the critical point:
\begin{equation}
\alpha = \left.\frac{d  
\lambda(t)}{dt}\right|_{t=t_a}
\label{eq:alphafactor}
\end{equation}
As the system is annealed towards the critical point, the equilibration time increases. If the initial $\lambda$ is far away from the critical point, the system quickly relaxes to its equilibrium and can follow the annealing schedule in quasi-equilibrium or adiabatically.
There comes a point, $\hat{t}$, when the remaining annealing time, $t_a - \hat{t}$ is precisely equal to the relaxation time of the system, $\tau[\lambda(\hat{t})]$. At this point, any further change in $\lambda$ that increases the relaxation time results in the system not being able to equilibrate by the end of the annealing. The idea of KZM, then, is to find this ``freeze-out'' boundary $\hat{t}$ and to relate non-equilibrium correlation lengths to equilibrium correlations lengths via $\xi[\lambda(\hat{t})]$, which is then related to a density of topological defects via dimensional analysis. Based on this sketch, we proceed to find $\hat{t}$ by equating $|t_a - \hat{t}|$ to $\tau(\hat{\lambda})$ using Eq.~\ref{eq:taucorr} and Eq.~\ref{eq:linanneal}. We obtain:
\begin{equation}
\hat{t'} = \bigg(\tau_0 \alpha^{-z \nu}\bigg)^{\frac{1}{1 + z\nu}}
\end{equation}
where $\hat{t'} = t_A - \hat{t}$.  We then find $\lambda(\hat{t'})$ as:
\begin{equation}
    \lambda(\hat{t'})=\lambda_c(1- \tau'_0 \alpha^{\frac{1}{1+z\nu}})
\end{equation}
where $\tau'_0$ is a modified $\tau_0$ with unimportant constants that do not affect the exponents. We then obtain the correlation lengths at freeze-out from: 
\begin{equation}
    \xi[\lambda(\hat{t})]= \epsilon'_0 \,\alpha^{^\frac{-\nu}{1+z\nu}}
\end{equation}
where $\epsilon'_0$ is a modified $\epsilon_0$ with unimportant constants. Finally, $\xi$ is related to the density of topological defects through dimensional analysis of defect dimension $d$ and system dimension $D$: 
\begin{equation}
\frac{\xi^d}{\xi^D} = \frac{1}{\epsilon'^{(D-d)}_0}\,\alpha^{^{\left(\frac{\nu}{1+z\nu}\right)(D-d)}}
\label{eq:kzm}
\end{equation}
Eq.~\ref{eq:kzm} predicts a power-law relationship for the number of ``defects''  as  a function of quench ``velocity'', $\alpha$ in terms of equilibrium critical exponents. Defects in the context of Ising models are then relatable to measurable quantities such as kinks~\cite{bando2020probing} and to the residual energy with similar power-law scaling laws. In this paper, we do not attempt to reproduce critical exponents for c-bits or p-bits, but simply observe the power-law dependence of c-bits and p-bits with respect to $\alpha$ in 2D ferro-Ising and 3D spin glass models.

\subsection{Appendix 2}
\label{sec:appendix2}

\begin{algorithm}
\caption{Adaptive Parallel Tempering Preprocessing with p-bits}\label{alg:preprocess}
    \KwIn{energy variance tolerance, number of chains, number of sweeps, number of samples, constant parameter $\alpha$}
    \KwOut{problem-specific $\beta$ schedule for parallel tempering}

    $i \leftarrow 1$\;
    $\beta_1 \leftarrow 0.5$\;
    Initialize chains to random spins\;

    \While{energy variance $\sigma_E$ is greater than tolerance}{
        \For{each chain in parallel}{
            Simulate p-bits using Eq.~\ref{eq:p-bit neuron} and~\ref{eq:synapse} for the specified number of Monte Carlo sweeps at constant inverse temperature $\beta_i$\;
            Calculate variance $\sigma_E$ of energies sampled at the end of the simulation\;
            Save the spin state $\{m\}$ of each chain as the initial condition for the next iteration\;
        }
      
        Calculate the mean energy variance across chains, $\langle \sigma_E \rangle$\;
    
        $\beta_{i+1} \leftarrow \beta_i + \frac{\alpha}{\langle \sigma_E \rangle}$\;
        $i \leftarrow i + 1$\;
    }
    
\end{algorithm}

We use the following parameters for APT preprocessing: 100 parallel chains, $10^4$ Monte Carlo sweeps, 1000 samples (taken at the end of $10^4$ sweeps), and $\alpha = 1.25$. Energy variance tolerance is equal to half of the smallest magnitude weight $J_{ij}$, and the initial value of $\beta$ is set to 0.5. We have used a similar preprocessing method in our previous work~\cite{nikhar_all--all_2023}.

\subsection{Appendix 3}
\label{sec:appendix3}

To derive our time-to-solution performance metric (Eq.~\ref{eq:TTS}), we start from:

\begin{align}
[1-p(Nt_a)]^k = 1 - 0.99
\label{eq:TTS_derive_1}
\end{align}

where $p(Nt_a)$ is the probability of finding the solution using $N$ system replicas each running for time $t_a$. The left-hand side of the equation represents the probability of failing $k$ times in a row, and the right-hand side is the complement of the desired success probability, 0.99. We solve for $k$ through algebraic manipulation. The time-to-solution is then:
\begin{align}
Nt_a \times k 
\label{eq:TTS_derive_2}
\end{align}

which yields our definition of time-to-solution (Eq.~\ref{eq:TTS}). This can be thought of as the total computational time of the algorithm, $Nt_a$, multiplied by the average number of attempts before success.

\section*{Acknowledgements}
We gratefully acknowledge discussions with Sanaaya Lakdawala and Navid Anjum Aadit,  Nikhil Shukla and Corentin Delacour for insights regarding Oscillator Ising Machines and Masoud Mohseni for insights on KZM. This work is partially supported by an Office of Naval Research Young Investigator Program grant, and a National Science Foundation CCF 2106260 grant. Use was made of computational facilities purchased with funds from the National Science Foundation (CNS-1725797) and administered by the Center for Scientific Computing (CSC). The CSC is supported by the California NanoSystems Institute and the Materials Research Science and Engineering Center (MRSEC; NSF DMR 2308708) at UC Santa Barbara.

\end{document}